\documentclass{article}
\usepackage{ijcai16}
\usepackage{times}
\usepackage{helvet}
\usepackage{courier}
\usepackage{booktabs}
\usepackage{epsfig}
\usepackage{amssymb}
\usepackage{amsmath}
\usepackage{amsfonts}
\usepackage{multirow}
\usepackage{array}
\usepackage{subcaption}
\usepackage{mathtools}
\usepackage{caption}
\usepackage{url}
\usepackage{epstopdf}

\newcommand{\model}{Sherlock} 

\newcolumntype{R}[1]{>{\raggedleft\arraybackslash}p{#1}}
\newcommand{\xhdr}[1]{\vspace{0mm}\noindent{{\bf #1}}}
\DeclareMathOperator*{\argmax}{arg\,max}

\title{\model: Sparse Hierarchical Embeddings for Visually-aware \\ One-class Collaborative Filtering}

\author{Ruining He, Chunbin Lin, Jianguo Wang, Julian McAuley \\ 
University of California, San Diego  \\
\{r4he, chunbinlin, csjgwang, jmcauley\}@cs.ucsd.edu}

\begin{document}

\maketitle
\begin{abstract}
Building successful recommender systems requires uncovering the underlying dimensions that describe the properties of items as well as users' preferences toward them. In domains like clothing recommendation, explaining users' preferences requires modeling the visual appearance of the items in question. This makes recommendation especially challenging, due to both the complexity and subtlety of people's `visual preferences,' as well as the scale and dimensionality of the data and features involved. Ultimately, a successful model should be capable of capturing considerable variance across different categories and styles, while still modeling the commonalities explained by `global' structures in order to combat the sparsity (e.g.~\emph{cold-start}), variability, and scale of real-world datasets. 
Here, we address these challenges by building such structures to model the visual dimensions across different product categories. With a novel hierarchical embedding architecture, our method accounts for both high-level (colorfulness, darkness, etc.) and subtle (e.g.~casualness) visual characteristics simultaneously.

\end{abstract}

\section{Introduction}
Identifying and understanding the dimensions of users' opinions is a key component of any modern recommender system. 
Low-dimensional representations form the basis for Matrix Factorization methods, which model the affinity of users and items via their low-dimensional embeddings.
In domains like clothing recommendation where the \emph{visual} factors are largely at play, it is crucial to identify the visual dimensions of people's opinions in order to model personalized preference most accurately. 


In such domains, the visual appearance of the items 
are crucial side-signals that facilitate \emph{cold-start} recommendation.
However, uncovering visual dimensions directly from the images is particularly challenging as it involves handling semantically complicated and high-dimensional visual signals. In real-world recommendation data, items often come from a rich category hierarchy and vary considerably.
Presumably there are dimensions that focus on globally-relevant visual rating facets (colorfulness,  certain specific pattern types, etc.) across different items, as well as 
subtler
dimensions that may have different semantics (in terms of the combination of raw features) for different categories. For instance, the combination of raw features that determine whether a coat is `formal' might be totally different from those for a watch. 
Collars, buttons, pockets, etc. are at play in the former case, while features that distinguish digital from analogue or different strap types are key in the latter. Thus extracting visual dimensions requires a deep understanding of the underlying fine-grained variances. 

Existing works address this task by learning a linear embedding layer (parameterized by a matrix) on top of the visual features extracted from product images with a pre-trained Deep Convolutional Neural Network (Deep CNN) \cite{VBPR,VisualSIGIR,updown}. This technique has seen success for recommendation tasks from visually-aware personalized ranking (especially in \emph{cold-start} scenarios), to discriminating the relationships between different items (complement, substitute, etc.). However, with a single embedding matrix shared by all items, such a model is only able to uncover `general' visual dimensions, but is limited in its ability to capture subtler dimensions. Therefore the task of modeling the visual dimensions of people's opinions is only partially solved.  

In this paper, we propose an efficient sparse hierarchical embedding method, called \emph{\model}, to uncover the visual dimensions of users' opinions on top of raw visual features (e.g.~Deep CNN features extracted from product images). With a flexible distributed architecture, \model~is scalable and allows us to simultaneously learn both general and subtle dimensions, captured by different layers on the category hierarchy. 
More specifically, our main contributions include:
\begin{itemize}
\item We propose a novel hierarchical embedding method for uncovering visual dimensions. Our method directly facilitates personalized ranking in the
one-class Collaborative Filtering setting, where only the `positive' responses (purchases, clicks, etc.) of users are available.
\item We quantitatively evaluate \model~on large e-comm-erce datasets: \emph{Amazon} Clothing Shoes \& Jewelry, etc. Experimental results demonstrate that \model{}~outperforms state-of-the-art methods in both \emph{warm-start} and \emph{cold-start} settings.
\item We visualize the visual dimensions uncovered by \model{}~and qualitatively analyze the differences with state-of-the-art methods. 
\end{itemize}


\section{Related Work}
\label{sec:related_work}
\xhdr{One-class Collaborative Filtering.} 
\cite{OCCF} introduced the concept of One-Class Collaborative Filtering (OCCF) to allow Collaborative Filtering (CF) methods, especially Matrix Factorization (MF), to 
model users' preferences in one-class scenarios where only \emph{positive} feedback (purchases, clicks, etc.) is observed. This line of work includes
\emph{point-wise} models \cite{WRMF,OCCF} which implicitly treat non-observed interactions as `negative' feedback. Such ideas were refined by \cite{BPR} to develop a Bayesian Personalized Ranking (BPR) framework to optimize \emph{pair-wise} rankings of positive versus non-observed feedback. BPR-MF combines the strengths of the BPR framework and the efficiency of MF and forms the basis of many state-of-the-art personalized OCCF methods (e.g.~\cite{GBPR,MRBPR,VBPR,ZhaoCIKMSocial}). 

\smallskip
\xhdr{Incorporating Side-Signals.} 
Despite their success, MF-based methods suffer from \emph{cold-start} issues due to the lack of observations to estimate the latent factors of new users and items. Making use of `side-signals' on top of MF approaches can provide auxiliary information in \emph{cold-start} settings while still maintaining MF's strengths. Such signals include the content of the items themselves, ranging
from the timbre and rhythm for music \cite{wang2014exploration,YahooMusic}, textual reviews that encode dimensions of opinions \cite{HiddenFactorsHiddenTopics,PhraselevelSentimentAnalysis,ling2014RMR}, or social relations \cite{ZhaoCIKMSocial,jamali2010matrix,GBPR,MRBPR}. 
Our work follows this line though we focus on a specific type of signal---visual features---which brings unique challenges when modeling the complex dimensions that determine users' and items' interactions.

\smallskip
\xhdr{Visually-aware Recommender Systems.}
The importance of using visual signals (i.e., product images) in recommendation scenarios has been stressed by previous works, e.g.~\cite{di2014picture,goswami2011study}. State-of-the-art visually-aware recommender systems make use of high-level visual features of product images extracted from (pre-trained) Deep CNNs on top of which a layer of parameterized transformation is learned to uncover those ``visual dimensions'' that predict users' actions (e.g.~purchases) most accurately. Modeling this transformation layer is non-trivial as it is crucial to maximally capture the wide variance across different categories of items without introducing prohibitively many parameters into the model. 

A recent line of work \cite{VBPR,updown,fashionista} makes use of a \emph{single} parameterized embedding matrix to transform items from the CNN feature space into a low-dimensional `visual space' whose dimensions (called `visual dimensions') are those that maximally explain the variance in users' decisions.
While a simple parameterization, this technique successfully captures the common features among different categories, e.g.~what characteristics make people consider a t-shirt or a shoe to be `colorful.' But it is limited in its ability to model
subtler variations,
since each item is modeled in terms of a low-dimensional, global embedding.
Here, we focus on building flexible hierarchical embedding structures that are capable of efficiently capturing both the 
common features and subtle variations across and within categories.

\smallskip
\xhdr{Taxonomy-aware Recommender Systems.}
There has been some effort to investigate `taxonomy-aware' recommendation models, including earlier works extending neighborhood-based methods (e.g.~\cite{ziegler2004taxonomy,weng2008exploiting}) and more recent endeavors to extend MF using either explicit (e.g.~\cite{mnih2012taxonomy}) or implicit (e.g.~\cite{zhang2014taxonomy,mnih2012learning,wang2015exploring}) taxonomies. 
This work is related to ours, though does not consider the visual appearance of products as we do here.


\section{The \model{}~Model} \label{sec:model}
We are interested in
learning
visually-aware preference predictors 
from a large corpus of implicit feedback, i.e., the `one-class' setting. Formally, let $\mathcal{U}$ and $\mathcal{I}$ be the set of users and items respectively. 
Each user $u \in \mathcal{U}$ has expressed \emph{positive} feedback about a set of items $\mathcal{I}_u^+$. Each item $i \in \mathcal{I}$ is associated with a visual feature vector $f_i \in \mathbf{R}^{F}$ extracted from a pre-trained Deep CNN. For each item we are also given a path on a category hierarchy from the root node to a leaf category $\mathcal{C}_i$ (e.g.~Clothing $\rightarrow$ Women's Clothing $\rightarrow$ Shoes). Our objective is to build an efficient visually-aware preference predictor based on which a \emph{personalized} ranking of the non-observed items (i.e., $\mathcal{I} \setminus \mathcal{I}_u^+$) is predicted for each user $u \in \mathcal{U}$. 

\subsection{The Basic Visually-aware Predictor}
A visually-aware personalized preference predictor can be built on top of MF \cite{VBPR}, which predicts the `affinity score' between user $u$ and item $i$, denoted by $\widehat{x}_{u,i}$, by their interactions on visual and non-visual dimensions simultaneously: 
\begin{equation} \label{eq:VBPR}
\widehat{x}_{u,i} = \underbrace{\langle \overbrace{\theta_u;\gamma_u}^{\text{user factors}}, \overbrace{\theta_i;\gamma_i}^{\text{item factors}} \rangle}_{\text{interactions between $u$ and $i$}} + \underbrace{\langle \vartheta, f_i \rangle}_{\text{visual bias of $i$}} + \beta_i,
\end{equation}
where user $u$'s preferences on visual and non-visual (latent) dimensions are represented by vectors $\theta_u \in \mathbf{R}^{K'}$ and $\gamma_u \in \mathbf{R}^K$ respectively. Correspondingly, $\theta_i \in \mathbf{R}^{K'}$ and $\gamma_i \in \mathbf{R}^K$ are two vectors encoding the visual and non-visual `properties' of item $i$. The affinity between $u$ and $i$ is then predicted by computing the inner product of the two concatenated vectors. Item $i$ is also associated with an offset composed of two parts: visual bias captured by the inner product of $\vartheta$ and $f_i$, and the (non-visual) residue $\beta_i$. 

\subsection{Modeling the Visual Dimensions}
The key to the above predictor is to model the \emph{visual} dimensions of users' opinions, which has seen success at tackling \emph{cold-start} issues especially for domains like clothing recommendation where visual factors are 
at play. Instead of standard dimensionality reduction techniques like PCA, prior works have found it 
more effective
to learn an embedding kernel (parameterized by a $K'\times F$ matrix $\mathbf{E}$) to linearly project items from the raw CNN feature space to a low-dimensional 
`visual space' \cite{VBPR,updown}. Critically, each dimension of the visual space encodes one visual facet that users consider;
these facets $\theta_i$ are thus modeled by $\theta_i = \mathbf{E}f_i$, i.e., all $\theta_i$'s are parameterized by a single matrix $\mathbf{E}$.

For each visual dimension $D$, the above technique computes the extent to which an item $i$ exhibits this particular visual rating facet (i.e., the `score' $\theta_i^{(D)}$) by the inner product
\begin{equation}
\theta_i^{(D)} = \langle \mathbf{E}^{(D)}, f_i \rangle,
\end{equation}
where $\mathbf{E}^{(D)}$ is the $D$-th row of $\mathbf{E}$. 
Despite its simple form, the embedding layer is able to uncover visual dimensions that capture `global' features
(e.g.~colorfulness, a specific type of pattern) across different items. However, in settings where items are organized 
in a rich category hierarchy, such a single-embedding method can only partially solve the task of modeling visual preferences as
dimensions that capture the subtle  
and wide variances across different categories are ignored. For example, the 
combination of raw
features that determine whether a shirt is casual (such as a zipper and short sleeves) might be quite different from those for a shoe. Thus to extract the `scores' on a visual dimension indicating casualness, each category may require a different parameterization.

It is the above properties that motivate us to use \emph{category-dependent} structures to extract different visual embeddings for each category of item.

\subsection{Sparse Hierarchical Embeddings}
Variability is pervasive in real-world datasets; this is especially true in e-commerce scenarios where the items from different parts of a category hierarchy may vary considerably. 

A simple idea would be to learn a separate embedding kernel for each (sub-) category to maximally capture the fine-grained variance. However, this would introduce prohibitively many parameters and would thus be subject to overfitting. For instance, the \emph{Amazon} Women's Clothing dataset with which we experiment has over 100 categories which would require millions of parameters just to model the embeddings themselves. Our key insight is that the commonalities among nodes sharing common ancestors mean that there would be considerable redundancy in the above scheme, which can be exploited with efficient structures.
This inspires us to develop a distributed architecture---a flexible hierarchical embedding model---in order to simultaneously account for the commonalities and variances efficiently. 

\begin{figure}
\centering
\includegraphics[width=\columnwidth]{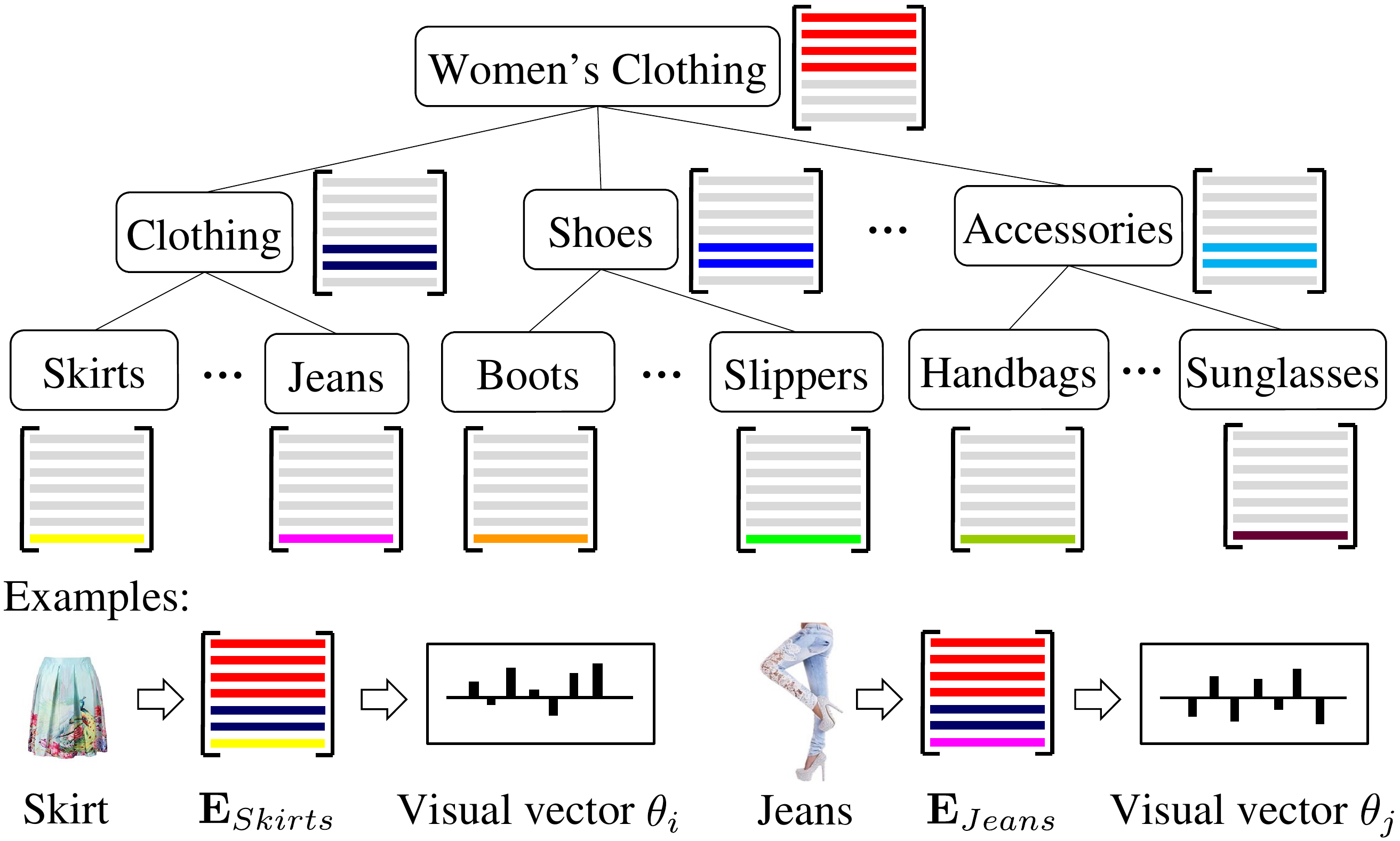}
\caption{Illustration of the high-level idea of \model{}. Each layer is associated with a \emph{segment} of the embedding matrix, which is instantiated at each node on this layer. The embedding matrix at each leaf node is a \textit{concatenation/stack} of the segments along the root-to-leaf path. (Here we demonstrate allocating $K'=7$ visual dimensions with a 4:2:1 split on the hierarchy $\Psi$.)}
\label{fig:idea}
\end{figure}

Note that each row in the $K' \times F$ embedding matrix $\mathbf{E}$ is used to extract a specific visual dimension. Let $\Psi$ denote the category hierarchy and $h_{\Psi}$ its height, the high-level idea is to partition the $K'$ row indices of the embedding matrix into $h_{\Psi}$ segments such that each segment is associated with a specific layer on $\Psi$. For each layer, the associated segment is instantiated with an \emph{independent} copy at each node which is used to extract the corresponding visual dimensions for all items in the subtree rooted with the node. 

Figure \ref{fig:idea} demonstrates a concrete example of extracting $K'=7$ visual dimensions for the \emph{Amazon} Women's Clothing dataset. Here we show the top 3 layers of the hierarchy (i.e., $h_{\Psi} = 3$); handling deep and complex hierarchies will be discussed later.
In this example, layers 0, 1, and 2 are associated with segment $S_0=[0,1,2,3]$, $S_1=[4,5]$, and $S_2=[6]$ respectively. The $4 \times F$ matrix associated with the root node will be used to extract the first 4 visual dimensions (i.e., $0$, $1$, $2$, $3$) for all items in the dataset. While the $2 \times F$ matrices attached to the second-level nodes (i.e.~Clothing, Shoes, etc.) are used to extract visual dimensions $4$ and $5$ for items in the corresponding subtrees. Finally, the $1 \times F$ matrices on the bottom layer are used to extract dimension $6$ for each of the fine-grained categories (i.e., Skirts, Jeans, Boots, etc.). 

Finally, each leaf node of $\Psi$ is associated with a full $K' \times F$ embedding matrix, which is the concatenation/stacking of all the segment instances along the root-to-leaf path.
Formally, let $(\mathcal{C}_i^{(1)}$, \ldots , $\mathcal{C}_i^{(h_{\Psi})})$ be the root-to-leaf path of item $i$ on $\Psi$, and $S_i^{(l)}$ the segment of the embedding matrix associated with node $C_i^{(l)}$ on layer $l$. The full embedding matrix associated with the leaf node $\mathcal{C}_i^{(h_{\Psi})}$ is:
$$
\mathbf{E}_{C_i^{(h_{\Psi})}} = 
\begin{bmatrix} 
\text{\textbf{-----}}~~S_i^{(1)}~~\text{\textbf{-----}} \\ 
\text{\textbf{-----}}~~S_i^{(2)}~~\text{\textbf{-----}} \\
\cdots \\
\text{\textbf{-----}}~~S_i^{(h_{\Psi})}\text{\textbf{-----}} \\
\end{bmatrix}_{K' \times F}.
$$

Then the `visual properties' of item $i$ (i.e., $\theta_i$) are computed by 
\begin{equation}
\theta_i = \mathbf{E}_{C_i^{(h_{\Psi})}}f_i.
\end{equation}

For example, in Figure \ref{fig:idea} the embedding matrix for the `Skirts' node ($\mathbf{E}_{Skirts}$) consists of 4 rows inherited from the root, 2 rows from its parent `Clothing,' and 1 row of its own. Each skirt item $i$ in the dataset will be projected to the visual space by $\mathbf{E}_{\mathit{Skirts}}f_i$. 

Note that parameters attached to the ancestor are shared by all its descendants and capture more globally-relevant visual dimensions (e.g.~colorfulness), while the more fine-grained dimensions (e.g.~a specific style type) are captured by descendants on lower levels. This `sparse' structure enables \model~to simultaneously capture multiple levels of dimensions with different semantics at a low cost. 

The distributed architecture of \model~allows tuning the partition schemes of the $K'$ visual dimensions to trade off the expressive power and the amount of free parameters introduced. Allocating more dimensions to higher layers can reduce the number of parameters, and the opposite will enhance the ability to capture more fine-grained characteristics. Note that \model~can capture the single-embedding model as a special case if it allocates all $K'$ dimensions to the root node, i.e., use a $K':0:\ldots :0$ split.

\xhdr{Imbalanced Hierarchies.} 
\model~is able to handle imbalanced category hierarchies. 
Let $\hbar$ be the length of the shortest path on an imbalanced hierarchy. \model~only assigns segments to layers higher than $\hbar$, using the same scheme as described before. In this way, the unbalanced hierarchy can be seen as being `reduced' to a balanced one. Note that (1) real-world category hierarchies (such as \emph{Amazon}'s catalogs) usually are not severely imbalanced, and (2) experimentally \model{} does not need to go deep down the hierarchy to perform very well (as we show later), presumably because data get increasingly sparser when diving deeper.

\subsection{Learning the Model}

Bayesian Personalized Ranking (BPR) is a state-of-the-art framework to directly optimize the personalized ranking $>_{u}$ for each user $u$ \cite{BPR}. Assuming independence of users and items, it optimizes the maximum a posterior (MAP) estimator of the model parameters $\Theta$:
\begin{equation}
\argmax_{\Theta} = \ln \prod_{u \in \mathcal{U}} \prod_{i \in \mathcal{I}_u^+} \prod_{j \notin \mathcal{I}_u^+} p(i >_{u} j | \Theta) ~ p(\Theta),
\end{equation}
where the \emph{pairwise} ranking between a positive ($i$) and a non-positive ($j$) item $p(i >_{u} j | \Theta)$ is estimated by a logistic function $\sigma(\widehat{x}_{u,i} - \widehat{x}_{u,j})$. 

Let $\mathbf{E}^*$ denote the set of embedding parameters associated with the hierarchy $\Psi$, then the full set of parameters of \model{}~becomes $\Theta = \{\beta_i, \gamma_i, \gamma_u, \theta_u, \vartheta, \mathbf{E}^*\}$. Using stochastic gradient ascent, first we uniformly sample a user $u$ as well as a positive/non-positive pair $(i,j)$, and then the learning procedure updates parameters as follows:
\begin{equation}
\Theta \leftarrow \Theta + \alpha \cdot (\sigma(\widehat{x}_{u,j} - \widehat{x}_{u,i}) \frac{\partial (\widehat{x}_{u,i} - \widehat{x}_{u,j})}{\partial \Theta} - \lambda_{\Theta}\Theta ), 
\end{equation}
where $\alpha$ represents the learning rate and $\lambda_{\Theta}$ is a regularization hyperparameter tuned with a held-out validation set.

\xhdr{Complexity Analysis.} Here we mainly compare \model{}~against the single-embedding model. For a sampled triplet $(u,i,j)$, both methods require $O(K'\times F + K)$ to compute $\widehat{x}_{u,i} - \widehat{x}_{u,j}$. Next, computing the partial derivatives for $\beta_i$, $\gamma_i$, $\gamma_u$, $\theta_u$, and $\vartheta$ take $O(1)$, $O(K)$, $O(K)$, $O(K')$, and $O(F)$ respectively. Note that both models take $O(K' \times F)$ for updating the embedding parameters due to the amount of parameters \emph{involved}. In summary, \model{}~as well as the single-embedding model take $O(K' \times F + K)$ for updating a sampled triple.

\section{Experiments}
\label{sec:experiments}
In this section, we conduct experiments on real-world data-sets to evaluate the performance of \model{}~and visualize the hierarchical visual dimensions it uncovers. 

\begin{table}
\centering
\renewcommand{\tabcolsep}{1.4mm}
\caption{Dataset statistics} 
\begin{tabular}{lrrrr} \toprule
Dataset              &\#users ($|\mathcal{U}|$) & \#items ($|\mathcal{I}|$)   &\#feedback  \\\midrule
Men's Clothing       & 1,071,567  & 220,478   & 1,554,834  \\
Women's Clothing     & 1,525,979  & 465,318   & 2,634,336  \\
Full Clothing        & 2,948,860  & 1,057,034  & 5,338,520  \\\bottomrule
 \end{tabular}
\label{table:dataset}
\end{table}

\subsection{Datasets}
We are interested in evaluating our method on the largest
datasets available. To this end, we adopted the \emph{Amazon} dataset introduced by \cite{VisualSIGIR}. In particular, we consider the Clothing Shoes \& Jewelry dataset, since (1) clothing recommendation is a
task where visual signals are at play, and (2) visual features have proven to be highly successful at addressing recommendation tasks
on this dataset. Thus it is a natural choice for comparing our method against these previous models. Additionally, we also evaluate all methods on its two largest subcategories---Men's and Women's Clothing \& Accessories. 
Statistics of these three datasets are shown in Table~\ref{table:dataset}. For simplicity, in this paper we denote them as Men's Clothing, Women's Clothing, and Full Clothing respectively. 

\xhdr{Category Tree.} There is a category tree associated with each of our clothing datasets. Figure \ref{fig:idea} demonstrates part of the hierarchy associated with Women's Clothing. On this hierarchy, we have 10 second-level categories (Clothing, Shoes, Watches, Jewelry, etc.), and 106 third-level categories (e.g.~Jeans, Pants, Skirts, etc. under the Clothing category).

\xhdr{Visual Features.} Following the setting from previous works \cite{VBPR,updown,VisualSIGIR}, we employ the Deep CNN visual features extracted from the Caffe reference model \cite{Caffe}, which implements the architecture comprising 5 convolutional layers and 3 fully-connected layers \cite{DeepCNNArchitecture}. See \cite{VisualSIGIR} for more details.  
In this experiment, we take the output of the second fully-connected layer FC7 to obtain a visual feature vector $f_i$ of $F=4096$ dimensions.

\subsection{Evaluation Protocol}
Here we mainly follow the leave-one-out protocol used by \cite{VBPR,BPR}, i.e., for each user $u$ we randomly sample a positive item $\mathcal{V}_u \in \mathcal{I}_u^+$ for validation and another $\mathcal{T}_u \in \mathcal{I}_u^+$ for testing. The remaining data is then 
used for training. 
Finally, all methods are evaluated in terms of the quality of the predicted personalized ranking 
using the average AUC (\emph{Area Under the ROC curve}) metric:
\begin{equation}
\mathit{AUC} =  \frac{1}{|\mathcal{U}|}  \sum_{u\in\mathcal{U}}   \frac{1}{|\mathcal{I}\setminus\mathcal{I}_u^+|}   \sum_{j \in \mathcal{I} \setminus \mathcal{I}_u^+}  \mathbf{1} (\widehat{x}_{u,\mathcal{T}_u} > \widehat{x}_{u,j}),
\end{equation}
where $\mathbf{1}(\cdot)$ is the indicator function and the evaluation goes through the pair set of each user $u$.

All our experiments were conducted on a single desktop machine with 4 cores (3.4GHz) and 32GB memory. In all cases, we use grid-search to tune the regularization hyperparameters to obtain the best performance on the validation set and report the corresponding performance on the test set. 

\subsection{Baselines}
We mainly compare our approach against BPR-based ranking methods, which are known to have state-of-the-art performance for implicit feedback dataset. The baselines we include for evaluation are:
\begin{itemize}
\item \textbf{Random (RAND):} This baseline ranks all items randomly for each user $u$.
\item \textbf{BPR-MF:} Introduced by \cite{BPR}, this baseline is the state-of-the-art method for personalized ranking in OCCF settings. It takes the standard Matrix Factorization \cite{Handbook} as the preference predictor.
\item \textbf{Visual-BPR (VBPR):} Built upon the BPR-MF model, it uncovers visual dimensions on top of Deep CNN features with a single embedding matrix \cite{VBPR}. This is a state-of-the-art visually-aware model for the task though. 
\item \textbf{VBPR-C:} This baseline makes use of category tree to extend VBPR by associating a bias term to each fine-grained category on the hierarchy.
\end{itemize}

In summary, these baselines are designed to demonstrate (1) the strength of state-of-the-art visually-\emph{unaware} method on our datasets (i.e., BPR-MF), (2) the effect of using the visual signals and modeling those visual dimensions (i.e., VBPR), and (3) the improvement of using the category tree signal to improve VBPR in a relatively straightforward manner (i.e., VBPR-C).

\subsection{Performance and Quantitative Analysis}
For each dataset, we evaluate all methods with two settings: \emph{Warm-start} and \emph{Cold-start}. The former focuses on measuring the overall ranking performance, while the latter the capability to recommend \emph{cold-start} items in the system. Following the protocol used by \cite{VBPR}, the \emph{Warm-start} evaluation is implemented by computing the average AUC on the full test set, while \emph{Cold-start} evaluates by only keeping the \emph{cold} items in the test set, i.e., those that appeared fewer than five times in the training set.

For fair comparisons, all BPR-based methods use 20 rating dimensions.\footnote{We experimented with more dimensions but only got negligible improvements for all BPR-based methods.}
For simplicity, VBPR, VBPR-C, and \model{}~all adopt a 10:10 split, i.e., $K=10$ latent plus $K'=10$ visual dimensions. To evaluate the performance of \model, we use three different allocation schemes for the 10 visual dimensions, denoted by (e1), (e2), and (e3) respectively. Table \ref{tb:schemes} summarizes the detailed dimension allocation of each scheme for each of the three datasets. For Men's and Women's Clothing, all three schemes allocate the 10 visual dimensions to the top 3 layers on the category hierarchy. In contrast, visual dimensions are allocated to the top 4 layers for Full Clothing as it is larger in terms of both size and variance.

\begin{table}
\centering
\renewcommand{\tabcolsep}{4mm}
\caption{Different evaluation schemes to allocate 10 visual dimensions.\protect\footnotemark{} 
More visual dimensions are distributed to \emph{lower} layers on the hierarchy in the order of (e1) $\Rightarrow$ (e2) $\Rightarrow$ (e3).} 
\begin{tabular}{cccc} \toprule
Scheme      & Full Clothing & Men's   &Women's \\\midrule
(e1)        & 6:2:1:1       & 7:2:1            & 7:2:1  \\
(e2)        & 4:3:2:1       & 5:3:2            & 5:3:2  \\
(e3)        & 2:3:3:2       & 3:4:3            & 3:4:3  \\\bottomrule
\end{tabular}
\label{tb:schemes}
\end{table}
\footnotetext{Numbers separated by colons denote the corresponding amount of dimensions allocated to each layer (from top to bottom).}

Note that comparison between \model~and VBPR using the same amount of embedding parameters would be unfair for VBPR, as in that case VBPR would have to use a larger value of $K'$ and would thus need to learn more parameters for $\theta_u$ and would be subject to overfitting.

Table~\ref{table:auc} shows the average AUC of all methods achieved on our three datasets. The main findings from this table can be summarized as follows:

\begin{table*}
\renewcommand{\tabcolsep}{3pt}
\centering
\caption{AUC on the test set (total number of dimensions is set to 20). (e1), (e2) and (e3) correspond to the three evaluation schemes in Table \ref{tb:schemes}. The best performing method in each case is boldfaced.}
\begin{tabular}{llccccccccccccccccc} \toprule
\multirow{2}{*}{Dataset}  &\multirow{2}{*}{Setting}   &(a)    &(b)    &(c)     &(d)    &(e1)   &(e2)   &(e3)   &\multicolumn{2}{c}{improvement} \\ 
                          &                           &RAND   &BPR-MF  &VBPR   &VBPR-C  &\model &\model &\model & e vs. c  & e vs. best \\ \midrule

\multirow{2}{*}{Full Clothing}    &\emph{Warm-start}  &0.5012 &0.6134 &0.7541  &0.7552 &0.7817 &0.7842 &\textbf{0.7868}  &4.3\%  &4.2\%  \\
                                  &\emph{Cold-start}  &0.4973 &0.5026 &0.6996  &0.7002 &0.7272 &\textbf{0.7400} &0.7375  &5.8\%  &5.7\%  \\[4pt]
                                 
\multirow{2}{*}{Men's Clothing}   &\emph{Warm-start}  &0.4967 &0.5937 &0.7156  &0.7161 &0.7297 &0.7341 &\textbf{0.7364}  &2.9\%  &2.8\%  \\
                                  &\emph{Cold-start}  &0.4993 &0.5136 &0.6599  &0.6606 &0.6774 &0.6861 &\textbf{0.6889}  &4.4\%  &4.3\%  \\[4pt]
                                 
\multirow{2}{*}{Women's Clothing} &\emph{Warm-start}  &0.4989 &0.5960 &0.7336  &0.7339 &0.7464 &0.7476 &\textbf{0.7519}  &2.5\%  &2.5\%  \\
                                  &\emph{Cold-start}  &0.4992 &0.4899 &0.6725  &0.6726 &0.6910 &0.6960 &\textbf{0.7008}  &4.2\%  &4.2\%  \\ \bottomrule
\end{tabular}
\label{table:auc}
\end{table*}

\begin{enumerate}
\item Visually-aware methods (VBPR, VBPR-C, and \model{}) outperform BPR-MF significantly, which proves the usefulness of visual signals and the efficacy of the Deep CNN features.
\item VBPR-C does not significantly outperform
VBPR, presumably because the category signals are already (at least partially) encoded by those visual features. This suggests that improving VBPR requires more creative ways to leverage such signals.
\item \model{} predicts most accurately amongst all methods in all cases (up to 5.7\%) especially for datasets with larger variance (i.e., Full Clothing) and in \emph{cold-start} settings, which shows the effectiveness of using our novel hierarchical embedding architecture.
\item As we `offload' more visual dimensions to lower layers on the hierarchy (i.e., (e1) $\Rightarrow$ (e2) $\Rightarrow$ (e3)), performance becomes increasingly better. This indicates the stability of \model{} as well as the benefits from capturing category-specific semantics.
\end{enumerate}

\subsection{Training Efficiency}
\begin{figure}
\centering
\includegraphics[width=\columnwidth]{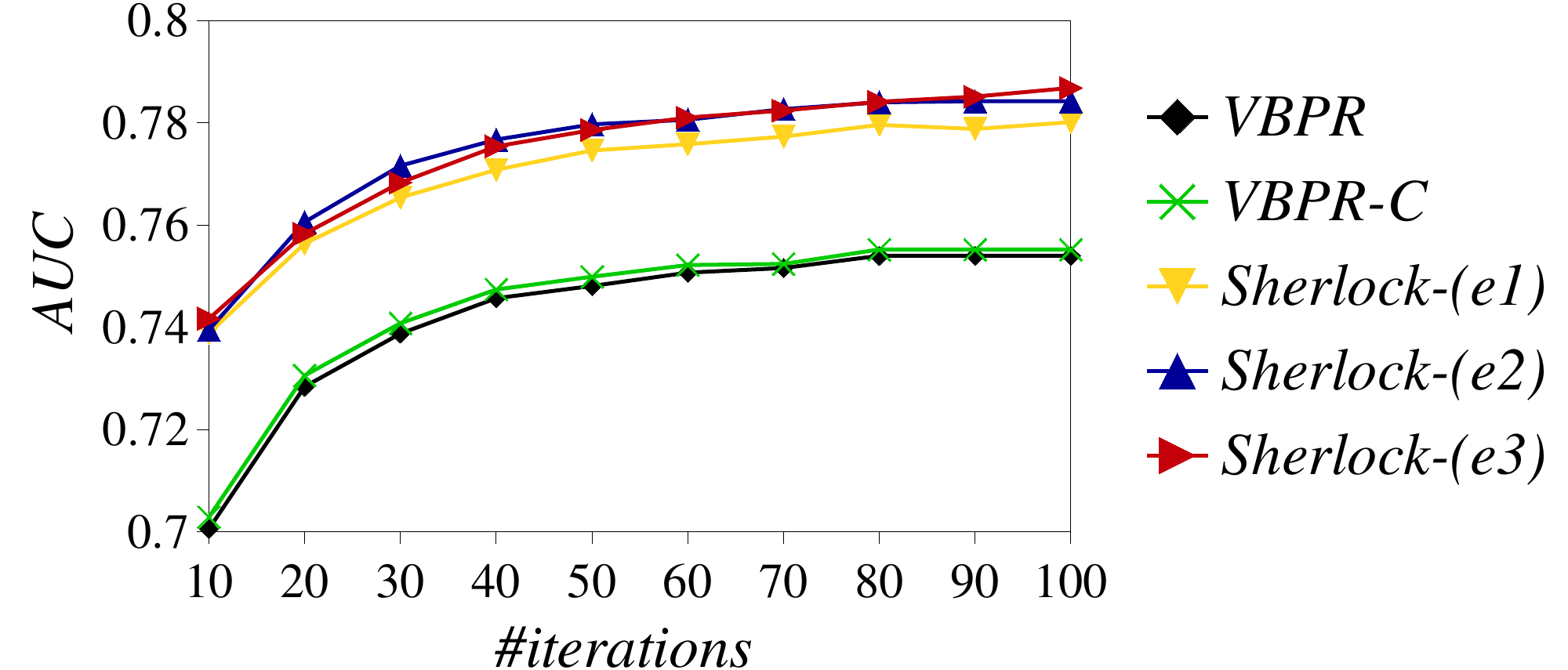}\\
\caption{Comparison of visually-aware methods with number of training iterations on Full Clothing dataset.}
\label{fig:curve}
\end{figure}

Each iteration of the stochastic gradient descent procedure consists of sampling training triples $(u,i,j)$ with the size of the training corpus. In Figure \ref{fig:curve} we demonstrate the accuracy on the test set (\emph{Warm-start} setting) of all visually-aware methods as the number of training iterations increases.
As shown by the figure, \model{} with three allocation schemes can all converge reasonably fast. 
As analyzed in Section \ref{sec:model}, our methods share the same time complexity with VBPR, and experimentally both are able to finish learning the parameters within a couple of hours.

\subsection{Visualization of Visual Dimensions}
Next we qualitatively demonstrate the visual dimensions revealed by \model. To achieve this goal, we take our model trained on Men's Clothing (10 visual dimensions with a split of 5:3:2\footnote{$D_0, D_1 \ldots, D_4$ are on the top-level, $D_5, D_6, D_7$ are on the second level, and $D_8, D_9$ are on the bottom-level.}) and rank all items on each visual dimension $D$ as follows:
$\argmax_{i} \theta_i^{(D)},$
i.e., to find items that maximally exhibit each dimension. 

\begin{figure}[h!]
\centering
\includegraphics[width=\columnwidth]{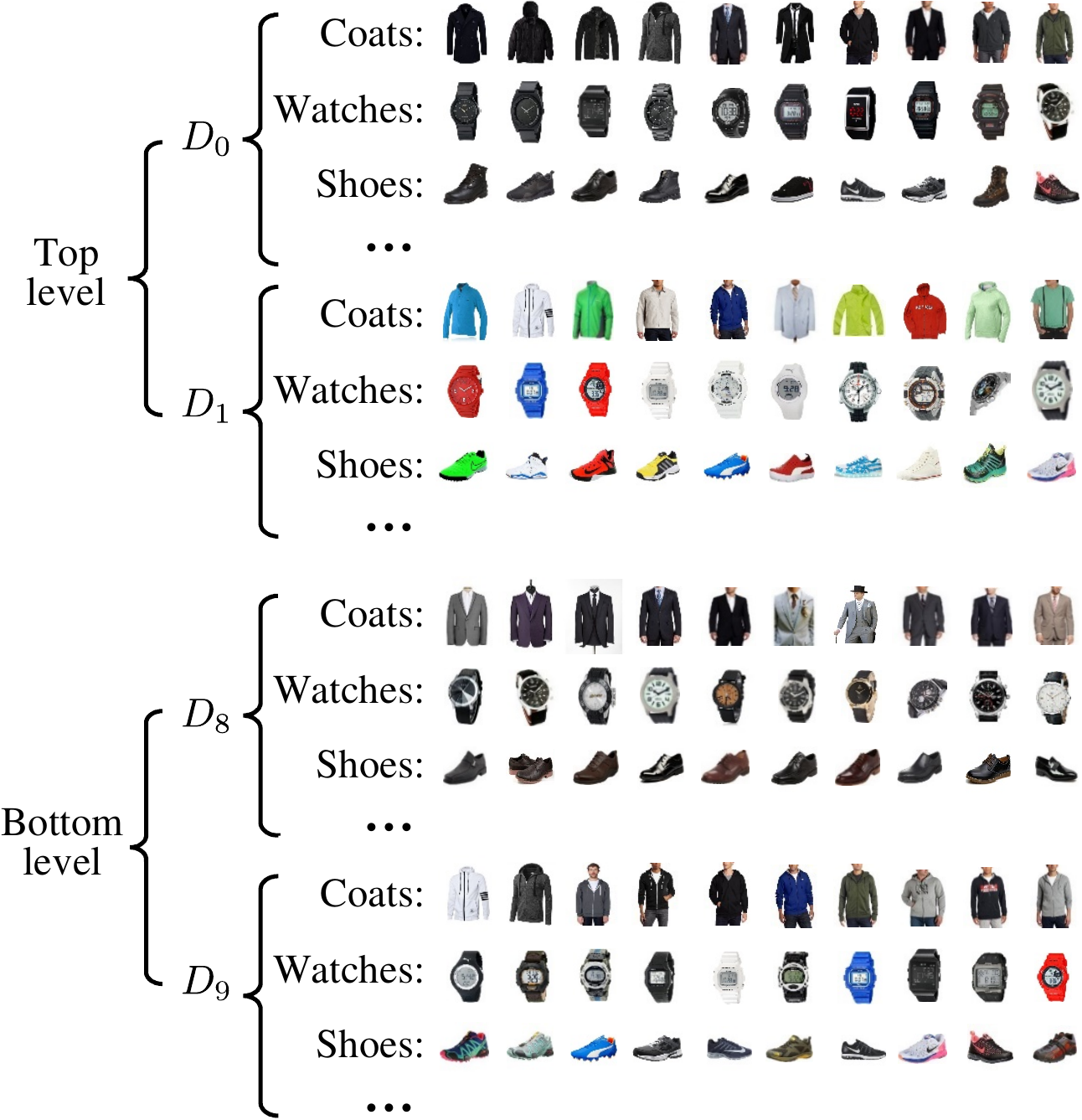}
\caption{Demonstration of four visual dimensions uncovered by \model~on Men's Clothing. Top-level dimensions $D_0$ and $D_1$ appear to capture `general' notions (darkness and brightness); while bottom-level dimensions $D_8$ and $D_9$ capture subtler notions (business and casualness), which requires the ability to model category-specific semantics.}
\label{fig:dims}
\end{figure}

In Figure \ref{fig:dims} we demonstrate four example dimensions, two associated with the top layer ($D_0, D_1$), and the rest the bottom layer ($D_8, D_9$). For each of these dimensions, we show the top-ranked items in a 
few of the bottom-level categories---Coats, Watches, and Shoes.
We make the following observations from Figure~\ref{fig:dims}.

\model~produces meaningful visual dimensions, each of which seems to capture a human notion. For example, dimensions $D_0$ and $D_8$ capture the notions of `dark' and `formality' respectively, though items are from different categories and vary considerably in aspects like shape and category-specific features.

As expected, \model~captures both globally-relevant and subtle dimensions. Top-level dimensions $D_0$ and $D_1$ are capturing the notions of darkness and brightness respectively, features which are applicable to all categories. While the bottom-level dimensions $D_8$ and $D_9$ are capturing the subtler notions of formality and casualness,
which requires a deep understanding of category-specific semantics. For example, to extract the `business' dimension ($D_8$), attributes like collar and button are at play for the Coats category, while features that distinguish digital from analogue or strap types are key for the Watches category.

Note the single-embedding model only uncovers dimensions like those captured by the top-level on our hierarchy.
The strength of \model~in exploiting the commonalities and variances of items and uncovering dimensions makes it a successful method in addressing the recommendation task. 

\section{Conclusion}
Uncovering rating dimensions and modeling user-item interactions upon them are key to building a successful recommender system. In this paper, we proposed a sparse hierarchical embedding method, \emph{\model}, that simultaneously reveals globally-relevant and subtle visual dimensions efficiently. We evaluated \model~for personalized ranking tasks in the one-class setting and found it to significantly outperform state-of-the-art methods on real-world datasets. 

\bibliographystyle{named}
\bibliography{visual}

\begin{thebibliography}{}

\bibitem[\protect\citeauthoryear{Di \bgroup \em et al.\egroup
  }{2014}]{di2014picture}
Wei Di, Neel Sundaresan, Robinson Piramuthu, and Anurag Bhardwaj.
\newblock Is a picture really worth a thousand words?:-on the role of images in
  e-commerce.
\newblock In {\em WSDM}, pages 633--642, 2014.

\bibitem[\protect\citeauthoryear{Goswami \bgroup \em et al.\egroup
  }{2011}]{goswami2011study}
Anjan Goswami, Naren Chittar, and Chung~H Sung.
\newblock A study on the impact of product images on user clicks for online
  shopping.
\newblock In {\em WWW}, pages 45--46, 2011.

\bibitem[\protect\citeauthoryear{He and McAuley}{2016a}]{updown}
Ruining He and Julian McAuley.
\newblock Ups and downs: Modeling the visual evolution of fashion trends with
  one-class collaborative filtering.
\newblock In {\em WWW}, 2016.

\bibitem[\protect\citeauthoryear{He and McAuley}{2016b}]{VBPR}
Ruining He and Julian McAuley.
\newblock {VBPR:} visual bayesian personalized ranking from implicit feedback.
\newblock In {\em AAAI}, 2016.

\bibitem[\protect\citeauthoryear{He \bgroup \em et al.\egroup
  }{2016}]{fashionista}
Ruining He, Chunbin Lin, and Julian McAuley.
\newblock Fashionista: A fashion-aware graphical system for exploring visually
  similar items.
\newblock In {\em WWW}, 2016.

\bibitem[\protect\citeauthoryear{Hu \bgroup \em et al.\egroup }{2008}]{WRMF}
Yifan Hu, Yehuda Koren, and Chris Volinsky.
\newblock Collaborative filtering for implicit feedback datasets.
\newblock In {\em ICDM}, pages 263--272, 2008.

\bibitem[\protect\citeauthoryear{Jamali and Ester}{2010}]{jamali2010matrix}
Mohsen Jamali and Martin Ester.
\newblock A matrix factorization technique with trust propagation for
  recommendation in social networks.
\newblock In {\em RecSys}, pages 135--142, 2010.

\bibitem[\protect\citeauthoryear{Jia \bgroup \em et al.\egroup }{2014}]{Caffe}
Yangqing Jia, Evan Shelhamer, Jeff Donahue, Sergey Karayev, Jonathan Long, Ross
  Girshick, Sergio Guadarrama, and Trevor Darrell.
\newblock Caffe: Convolutional architecture for fast feature embedding.
\newblock In {\em MM}, pages 675--678, 2014.

\bibitem[\protect\citeauthoryear{Koenigstein \bgroup \em et al.\egroup
  }{2011}]{YahooMusic}
Noam Koenigstein, Gideon Dror, and Yehuda Koren.
\newblock Yahoo! music recommendations: modeling music ratings with temporal
  dynamics and item taxonomy.
\newblock In {\em RecSys}, pages 165--172, 2011.

\bibitem[\protect\citeauthoryear{Krizhevsky \bgroup \em et al.\egroup
  }{2012}]{DeepCNNArchitecture}
Alex Krizhevsky, Ilya Sutskever, and Geoffrey~E Hinton.
\newblock Imagenet classification with deep convolutional neural networks.
\newblock In {\em NIPS}, pages 1097--1105, 2012.

\bibitem[\protect\citeauthoryear{Krohn-Grimberghe \bgroup \em et al.\egroup
  }{2012}]{MRBPR}
Artus Krohn-Grimberghe, Lucas Drumond, Christoph Freudenthaler, and Lars
  Schmidt-Thieme.
\newblock Multi-relational matrix factorization using bayesian personalized
  ranking for social network data.
\newblock In {\em WSDM}, pages 173--182, 2012.

\bibitem[\protect\citeauthoryear{Ling \bgroup \em et al.\egroup
  }{2014}]{ling2014RMR}
Guang Ling, Michael~R Lyu, and Irwin King.
\newblock Ratings meet reviews, a combined approach to recommend.
\newblock In {\em RecSys}, pages 105--112, 2014.

\bibitem[\protect\citeauthoryear{McAuley and
  Leskovec}{2013}]{HiddenFactorsHiddenTopics}
Julian McAuley and Jure Leskovec.
\newblock Hidden factors and hidden topics: understanding rating dimensions
  with review text.
\newblock In {\em RecSys}, pages 165--172, 2013.

\bibitem[\protect\citeauthoryear{McAuley \bgroup \em et al.\egroup
  }{2015}]{VisualSIGIR}
Julian McAuley, Christopher Targett, Qinfeng Shi, and Anton van~den Hengel.
\newblock Image-based recommendations on styles and substitutes.
\newblock In {\em SIGIR}, pages 43--52, 2015.

\bibitem[\protect\citeauthoryear{Mnih and Teh}{2012}]{mnih2012learning}
Andriy Mnih and Yee~W Teh.
\newblock Learning label trees for probabilistic modeling of implicit feedback.
\newblock In {\em NIPS}, pages 2816--2824, 2012.

\bibitem[\protect\citeauthoryear{Mnih}{2012}]{mnih2012taxonomy}
Andriy Mnih.
\newblock Taxonomy-informed latent factor models for implicit feedback.
\newblock In {\em KDD Cup}, pages 169--181, 2012.

\bibitem[\protect\citeauthoryear{Pan and Chen}{2013}]{GBPR}
Weike Pan and Li~Chen.
\newblock {GBPR}: group preference based bayesian personalized ranking for
  one-class collaborative filtering.
\newblock In {\em IJCAI}, pages 2691--2697, 2013.

\bibitem[\protect\citeauthoryear{Pan \bgroup \em et al.\egroup }{2008}]{OCCF}
Rong Pan, Yunhong Zhou, Bin Cao, Nathan~N Liu, Rajan Lukose, Martin Scholz, and
  Qiang Yang.
\newblock One-class collaborative filtering.
\newblock In {\em ICDM}, pages 502--511, 2008.

\bibitem[\protect\citeauthoryear{Rendle \bgroup \em et al.\egroup }{2009}]{BPR}
Steffen Rendle, Christoph Freudenthaler, Zeno Gantner, and Lars Schmidt-Thieme.
\newblock {BPR}: Bayesian personalized ranking from implicit feedback.
\newblock In {\em UAI}, pages 452--461, 2009.

\bibitem[\protect\citeauthoryear{Ricci \bgroup \em et al.\egroup
  }{2011}]{Handbook}
Francesco Ricci, Lior Rokach, Bracha Shapira, and Paul Kantor.
\newblock {\em Recommender systems handbook}.
\newblock Springer US, 2011.

\bibitem[\protect\citeauthoryear{Wang \bgroup \em et al.\egroup
  }{2014}]{wang2014exploration}
Xinxi Wang, Yi~Wang, David Hsu, and Ye~Wang.
\newblock Exploration in interactive personalized music recommendation: a
  reinforcement learning approach.
\newblock {\em ACM Transactions on Multimedia Computing, Communications, and
  Applications}, 11(1), 2014.

\bibitem[\protect\citeauthoryear{Wang \bgroup \em et al.\egroup
  }{2015}]{wang2015exploring}
Suhang Wang, Jiliang Tang, Yilin Wang, and Huan Liu.
\newblock Exploring implicit hierarchical structures for recommender systems.
\newblock In {\em IJCAI}, pages 1813--1819, 2015.

\bibitem[\protect\citeauthoryear{Weng \bgroup \em et al.\egroup
  }{2008}]{weng2008exploiting}
Li-Tung Weng, Yue Xu, Yuefeng Li, and Richi Nayak.
\newblock Exploiting item taxonomy for solving cold-start problem in
  recommendation making.
\newblock In {\em ICTAI}, pages 113--120, 2008.

\bibitem[\protect\citeauthoryear{Zhang \bgroup \em et al.\egroup
  }{2014a}]{PhraselevelSentimentAnalysis}
Yongfeng Zhang, Guokun Lai, Min Zhang, Yi~Zhang, Yiqun Liu, and Shaoping Ma.
\newblock Explicit factor models for explainable recommendation based on
  phrase-level sentiment analysis.
\newblock In {\em SIGIR}, pages 83--92, 2014.

\bibitem[\protect\citeauthoryear{Zhang \bgroup \em et al.\egroup
  }{2014b}]{zhang2014taxonomy}
Yuchen Zhang, Amr Ahmed, Vanja Josifovski, and Alexander Smola.
\newblock Taxonomy discovery for personalized recommendation.
\newblock In {\em WSDM}, pages 243--252, 2014.

\bibitem[\protect\citeauthoryear{Zhao \bgroup \em et al.\egroup
  }{2014}]{ZhaoCIKMSocial}
Tong Zhao, Julian McAuley, and Irwin King.
\newblock Leveraging social connections to improve personalized ranking for
  collaborative filtering.
\newblock In {\em CIKM}, pages 261--270, 2014.

\bibitem[\protect\citeauthoryear{Ziegler \bgroup \em et al.\egroup
  }{2004}]{ziegler2004taxonomy}
Cai-Nicolas Ziegler, Georg Lausen, and Lars Schmidt-Thieme.
\newblock Taxonomy-driven computation of product recommendations.
\newblock In {\em CIKM}, pages 406--415, 2004.

\end{thebibliography}

\end{document}